\documentclass[12pt]{article} 
\usepackage{amsmath,amssymb,mathrsfs,float,afterpage}
\usepackage[longnamesfirst]{natbib}
\usepackage{graphicx}
\usepackage{caption}
\usepackage{xcolor}
\begin{document}
\pagestyle{plain}
\parindent 0mm

\newcommand{\spc}[1]{\Pisymbol{cryst}{#1}}
\newcommand{\spctf}[1]{\begin{turn}{45} \Pisymbol{cryst}{#1}
\end{turn}} 
\newcommand{\spctn}[1]{\begin{turn}{90} \Pisymbol{cryst}{#1}
\end{turn}}

\def\bA{\mbox{\boldmath$A$}}
\def\bA{\mbox{\boldmath$A$}}
\def\ba{\mbox{\boldmath$a$}}
\def\bB{\mbox{\boldmath$B$}}
\def\bb{\mbox{\boldmath$b$}}
\def\bC{\mbox{\boldmath$C$}}
\def\bc{\mbox{\boldmath$c$}}
\def\bD{\mbox{\boldmath$D$}}
\def\bd{\mbox{\boldmath$d$}}
\def\bE{\mbox{\boldmath$E$}}
\def\be{\mbox{\boldmath$e$}}
\def\bF{\mbox{\boldmath$F$}}
\def\boldf{\mbox{\boldmath$f$}}
\def\bG{\mbox{\boldmath$G$}}
\def\bH{\mbox{\boldmath$H$}}
\def\bh{\mbox{\boldmath$h$}}
\def\bI{\mbox{\boldmath$I$}}
\def\bJ{\mbox{\boldmath$J$}}
\def\bK{\mbox{\boldmath$K$}}
\def\bk{\mbox{\boldmath$k$}}
\def\bL{\mbox{\boldmath$L$}}
\def\bM{\mbox{\boldmath$M$}}
\def\bN{\mbox{\boldmath$N$}}
\def\bn{\mbox{\boldmath$n$}}
\def\bP{\mbox{\boldmath$P$}}
\def\bp{\mbox{\boldmath$p$}}
\def\bq{\mbox{\boldmath$q$}}
\def\bR{\mbox{\boldmath$R$}}
\def\br{\mbox{\boldmath$r$}}
\def\bS{\mbox{\boldmath$S$}}
\def\bT{\mbox{\boldmath$T$}}
\def\bt{\mbox{\boldmath$t$}}
\def\bU{\mbox{\boldmath$U$}}
\def\bu{\mbox{\boldmath$u$}}
\def\bv{\mbox{\boldmath$v$}}
\def\bW{\mbox{\boldmath$W$}}
\def\bw{\mbox{\boldmath$w$}}
\def\bX{\mbox{\boldmath$X$}}
\def\bx{\mbox{\boldmath$x$}}
\def\by{\mbox{\boldmath$y$}}
\def\bZ{\mbox{\boldmath$Z$}}
\def\bz{\mbox{\boldmath$z$}}

\def\bbeta{\mbox{\boldmath$\beta$}}
\def\beps{\mbox{\boldmath$\epsilon$}}
\def\bSigma{\mbox{\boldmath $\Sigma$}}
\def\btau{\mbox{\boldmath $\tau$}}
\def\btheta{\mbox{\boldmath $\theta$}}
\def\blam{\mbox{\boldmath $\lambda$}}
\def\bphi{\mbox{\boldmath $\phi$}}
\def\bgam{\mbox{\boldmath $\gamma$}}

\def\cc{{\cal C}}
\def\calr{{\cal R}}
\def\cw{{\cal W}}
\def\cx{{\cal X}}
\def\cz{{\cal Z}}

\def\bzero{\mbox{\boldmath $0$}}
\def\b1{\mbox{\boldmath $1$}}
\def\bsim{\mbox{\boldmath $\sim$}}

\def \ni{\noindent}
\def \ds{\displaystyle}
\def \ul{\underline}
\def \fns{\footnotesize}
\def \ds{\displaystyle}

\def\be{\begin{equation}}
\def\ee{\end{equation}}

\newcommand{\hg}[2]{\mbox{}_{\scriptscriptstyle #1} F_{\scriptscriptstyle #2}}
\def \hlam{\hat{\lam}}
\def \hp{\hat{p}}
\newcommand{\overbar}[1]{\mkern 1.5mu\overline{\mkern-1.5mu#1\mkern-1.5mu}\mkern 1.5mu}

\def \mur{\frac{\mu}{\mu+r}}
\def \rmu{\frac{r}{\mu+r}}

\def \a{\alpha}
\def \b{\beta}
\def \s{\sigma}
\def \e{\epsilon}
\def \ul{\underline}
\def \t{\theta}
\def \ds{\displaystyle}
\def \d{\delta}
\def \g{\gamma}
\def \lam{\lambda}
\def \om{\omega}

\begin{center}
{\bf Exact Likelihoods for N-mixture models with Time-to-Detection Data}
\end{center}
\begin{center}
Linda M. Haines: 
{\small Department of Statistical Sciences,
University of Cape Town, \\Rondebosch 7700, South Africa.
{\small Email: {linda.haines@uct.ac.za}}}
\end{center}
\begin{center}
Res Altwegg:
{\small Centre for Statistics in Ecology, Environment and Conservation, \\Department of Statistical Sciences, University of Cape Town, \\
Rondebosch 7700, South Africa.
{\small Email: {res.altwegg@uct.ac.za}}}
\end{center}
\begin{center}
David L. Borchers:
{\small Centre for Research into Ecological and Environmental Modelling,\\
School of Mathematics and Statistics, The Observatory, Buchanan Gardens,\\
University of St Andrews,
Fife, KY16 9LZ, U.K.
{\small Email: {dlb@st-andrews.ac.uk}}}
\end{center}
\bigskip

\begin{abstract}
This paper is concerned with the formulation of $N$-mixture models for estimating the abundance and probability of detection of a species from binary response, count and time-to-detection data. A modelling framework, which encompasses time-to-first-detection within the context of detection/non-detection and time-to-each-detection and time-to-first-detection within the context of count data, is introduced. Two observation processes which depend on whether or not double counting is assumed to occur are also considered. The main focus of the paper is on the derivation of explicit forms for the likelihoods associated with each of the proposed models.  Closed-form expressions for the likelihoods associated with time-to-detection data are new and are developed from the theory of order statistics. A key finding of the study is that,  based on the assumption of no double counting, the likelihoods associated with times-to-detection together with count data are the product of the likelihood for the counts alone and a term which depends on the detection probability parameter. This result demonstrates that, in this case, recording times-to-detection could well improve precision in estimation over recording counts alone. In contrast, for the  double counting protocol with exponential arrival times, no information was found to be gained by recording times-to-detection in addition to the count data. An R package and an accompanying vignette are also introduced in order to complement the algebraic results and to demonstrate the use of the models in practice. 
\end{abstract}
\bigskip
\bigskip

{\bf 1. Introduction}
\label{sec:intro}

Models for the estimation of abundance and probability of detection of a species from surveys with binary response and count data recorded over a number of sites and occasions within an area of interest are well-documented and widely used \citep{kr:16}. 
The question immediately arises however as to whether or not recording the times-to-detection of the species in such surveys can, in some sense, improve the precision of the parameter estimates and, indeed, the stability and extent of the modelling process.  This question has been addressed within the context of occupancy models in a number of papers,  including the recent papers  by \cite{henry:20}, \cite{halstead:21} and \cite{reshwang:22}, but very much less often within the context of N-mixture models, for example in the  papers by \citet{martins:17} and \citet{strebel:21}.  The evidence presented thus far, specifically in relation  to an increase in the precision of the parameter estimates for models from time-to-detection data as an addition to binary response and count data, remains somewhat equivocal and more work, in both theory  and practice, is needed in order to clarify and, to some extent, to resolve the issue.

In the present paper, a broad-based modelling framework for the estimation of the abundance and probability of detection of a species based on binary responses, counts and time-to-detection data taken over a range of sites and occasions is introduced. The models proposed are, in essence, $N$-mixture models and the study focuses on the development of explicit expressions for the attendant likelihoods and on the use of the models in practice.  Five basic models are introduced, the detection-nondetection model of \citet{roylen:03} and the attendant time-to-first-detection model and the binomial $N$-mixture model of \citet{royle:04} which is extended in two ways to accommodate time-to-detection data. The derivations of the explicit likelihoods are intricate and lengthy and the formulae have been checked numerically against those evaluated by introducing an upper bound on the embedded infinite sum. The aim of the paper is to  provide a resource  which is available to researchers for modelling binary response, count and time-to-detection data as $N$-mixture models and for computing the requisite exact likelihoods.

The paper  is organised as follows.  The broad modelling framework and the five proposed models are presented in Section 2. In Section 3, explicit forms of the likelihood for the models of interest are developed, based on the premise that individuals, if detected more than once  on a single occasion, are only counted once. In accord with this assumption, the Bernoulli distribution is introduced for modelling the binary responses and the binomial distribution for modelling the counts.  A different model for the counts is introduced in Section 4, based on the premise that the same individual can be detected and counted more than once on a single occasion. Specifically, a homogeneous Poisson process is used to model the arrivals of an individual at a site \citep{garml:12}, in accord with a `double counting' setting \citep{lsbarker:18}.  An R package and an illustrative  vignette, are introduced in Section 5 in order to demonstrate the applicability of the models in practice. Some pointers for further research are presented in Section 6. 
\bigskip
\bigskip

{\bf 2. Preliminaries}
\label{sec:prelim}

Suppose that an ecologist would like to know the abundance, that is the number of individuals, of a particular species such as an animal or plant at $R$ sites and has decided to survey each of these sites $J$ times, which could be visiting each site only once $(J = 1)$ or multiple times $(J>1)$. The ecologist may consider different options for the survey protocol: 1) during each visit to each site, record whether or not the species is  detected; 2)  record the time until the species is detected,  that is the first individual of the species and note down the total time spent at the site if the species is not detected; 3) count all individuals of the species that are encountered during a visit over a fixed length of time; 4) in addition to counting individuals,  record the time to each detection; and 5)  count all individuals  encountered but only record time to the first detection. The ecologist may also  consider whether double-counting of the same individuals can be avoided or not. 
The flow chart given in Figure 1  provides a compact overview of the  models introduced in the present study and can be used by the ecologist to decide on a suitable survey protocol. The chart is also designed to be a valuable tool in navigating the formal developments which now follow, in particular the somewhat intricate derivations delineated in Section 3. 

\begin{center}
[FIGURE 1 ABOUT HERE]
\end{center}

The setting of interest can be stated more formally as follows. Suppose that a survey is to be conducted over an area of ecological interest with $R$ sites visited on each of $J$ occasions.  The number of individuals detected over an interval of time are to be recorded on each occasion, either as a binary $1/0$ variable, reflecting detection/non-detection, or as a count and, in addition, certain times-to-detection may be logged.
A general framework for building models for the survey is now developed. 
Two basic assumptions are made, first that the number of individuals at a site remains unchanged throughout the survey and second that individuals are detected independently of one another.

For site $i$ on occasion $j$, a binary or count response $y_{ij}, i=1, \ldots, R, j=1, \ldots, J$, is recorded. The number of individuals at a site $i$, denoted $n_i$, is unobserved and is taken to be an independent  draw from a discrete distribution with unknown parameter $\bphi$.  The pmf of $n_i$ is denoted compactly as $[n_i|\bphi]$. The observation process is driven by the time of arrival of an individual and the attendant continuous distribution  has unknown parameter $\btheta$.  However, the time-to-detection of an individual also depends on the search time, $T_{ij}$. Thus, a binary observation or a count recorded at site $i$ on occasion $j$  depends not only on the number of individuals at the site, $n_i$, but also on the unknown parameter $\btheta$ and on the search time $T_{ij}$, with the latter recorded and therefore known.  The pmf of $y_{ij}$ can thus be represented as $[y_{ij}|\btheta, n_i]$, with the search time $T_{ij}$ suppressed.

The basic form of the model is developed by summing over all possible values of $n_i$  for a given site $i$ and is expressed as follows
\begin{equation}
f(\by;\btheta,\bphi) = \prod_{i=1}^R\sum_{n_i={\kappa_i}}^{\infty} \Big ( \prod_{j=1}^J[y_{ij}|\btheta,n_i] \Big ) [n_i|\bphi],
\label{eq:genl}
\end{equation}
where $\by=(y_{11},\ldots,y_{R1},\ldots,y_{1J},\ldots,y_{RJ})$ is the vector of binary observations or counts, and $\kappa_i$ is the maximum recorded response, that is $\kappa_i=\underset{j=1,\ldots, J}{\max} ~y_{ij}$.
Models in which times-to-detection associated with each $y_{ij}$ are  also  considered. Thus, for a count $y_{ij}$, there may be an associated vector of times $\bt_{ij}=(t_{ij1},\ldots,t_{ijy_{ij}})$ recorded and for $y_{ij}$ binary, there may be a single such time $\bt_{ij}=t_{ij1}$. The general form of the model is then given by 
\begin{equation}
f(\by, \bt_{y};\btheta,\bphi) = \prod_{i=1}^R \sum_{n_i={\kappa_i}}^{\infty} \Big ( \prod_{j=1}^J \Big ( [y_{ij}|\btheta,n_i][\bt_{ij}|\btheta,y_{ij}] \Big )[n_i|\bphi],
\label{eq:genl.t}
\end{equation}
where $[\bt_{ij}|\btheta,y_{ij}]$ is the pdf of observed times given $y_{ij}$ and the unknown parameter vector $\btheta$ and $\bt_y$ represents a vector of times for a binary response or a vector formed from a ragged matrix for a count response.  When viewed as a function of the parameters $\btheta$ and $\bphi$ given $\by$ or  $\by$ and $\bt_{y}$, the formulae (\ref{eq:genl}) and (\ref{eq:genl.t}) are likelihood functions for the unknown parameters and maximum likelihood estimates can be obtained by maximising the likelihood with respect to the parameters.

Five  broad-based models are now introduced and are characterised by the nature of the observations. 
The responses  for each model are as follows:
\begin{itemize}
\itemsep0em
\item Binary: ~~~~Observed detection/non-detection.
\item BinaryT1: 
Time-to-first-detection.
\item Count: ~~~~~Total number of individuals detected.
\item CountT: ~~~Time-to-each-detection.
\item CountT1: ~~Time-to-first-detection and total number of individuals detected.
\end{itemize}
Within each model, a distinction is made  between those in which there is a single occasion, that is $J=1$, and those with multiple occasions, that is $J > 1$, by introducing notation  of the form Binary:S and Binary:M through CountT1:S and CountT1:M. 
The Binary and BinaryT1 models are varieties of detection/non-detection models, while the  Count and CountT models are versions of binomial $N$-mixture models, and the CountT1 models  are hybrid models. The  models and their interrelationships are neatly summarised in the flow chart in Figure 1. 
\bigskip
\bigskip

{\bf 3. Binomial Count Approach}
\label{sec:bincount}

The distribution of the unknown number of individuals at site $i$, that is $n_i$,  is taken  to be Poisson with parameter $\lam_i(\bphi)$, $i=1, \ldots, R$, and specifies the  mixing distribution.  The modelling framework itself, given  $n_i$, is based here  on the assumption that  an individual, if detected more than once on a given occasion, is only  counted once. In accord with this, the binary responses or counts  are taken to  follow a Bernoulli or binomial distribution and  the time of detection of an individual at site $i$ on occasion $j$ is taken to be exponentially distributed with parameter the rate of arrival of an individual per unit time, that is the hazard,  $h_{ij}(\btheta)$, $i=1, \ldots, R, j=1, \ldots, J$. For ease of presentation in the derivations which follow, the parameters are denoted  $\lam_i$ and $h_{ij}$, with the dependence on the unknown parameters $\bphi$ and $\btheta$ suppressed. Thus, for example, the probability of occurrence at site $i$ is given by $1-e^{-\lam_i}$ and the probability of detection of an individual at site $i$ on occasion $j$ by $1-e^{-h_{ij} T_{ij}}$, where $T_{ij}$ is the search time.
In addition, irrelevant factors, that is factors which only include observed responses, are usually omitted from the explicit expressions for the likelihoods. The formulae can however be easily amended if the Akaike information criterion (AIC) is required. The development of the models presented here follows that shown in the flow chart in Figure 1.
\bigskip

{\bf 3.1 Models for Detection/Non-detection}
\label{sec:M1}

The Binary:S and Binary:M models are  Royle-Nichols models with a  Poisson mixing distribution  \citep{roylen:03}. As a consequence, the likelihoods of the parameters for these models can be expressed in closed form by extending the results for the  Royle-Nichols model with  constant Poisson parameter and  constant probability of detection of an individual  derived in the paper by \cite{haines:16b}. 
\vspace{2.5mm}

\noindent{\sc Binary:S} 
 
Consider a site $i$ with a binary response $y_i, i=1, \ldots, R$. Then it follows from the assumption that the time of arrival of an individual is exponentially distributed with parameter $h_i$ that the probability of an individual not being detected over the search time $T_{i}$ is  $e^{-h_i T_i}$ and, since  individuals are detected independently of one another,  the probability that no individuals are detected over the  time $T_i$ is given by $P(y_i=0|n_i)= e^{-n_i h_i T_i}$, $i=1, \ldots, R$. The Poisson mixing distribution with parameter $\lam_i$ can now be introduced to yield the probability of no detections at the site as 
\begin{equation}
P(y_i = 0) = \sum_{n_i=0}^{\infty}  e^{-n_i h_i T_i} \frac{\lam_i^{n_i}  e^{\lam_i}}{n_i!} =\exp[-\lam_i (1 - e^{-h_i T_i})].
\label{proby0s}
\end{equation}
Clearly $P(y_i = 1) = 1-P(y_i = 0)$. Suppose now that the $R$ sites are arranged so that, without loss of generality, the species of interest is detected at the first $R_1$ sites and not detected at the remaining $R - R_1$ sites. Then, the likelihood of the parameters of the model Binary:S is given by
$$
L_{Binary:S}(\btheta, \bphi;\by)=  \prod_{i=1}^{R_1} \big\{ 1- \exp[-\lam_i (1 - e^{-h_i T_i})]  \big\}\prod_{i=R_1+1}^{R} \exp[-\lam_i (1 - e^{-h_i T_i})],
$$
where $\by=(y_1, \ldots, y_R)$.
Note  that, if the parameters $ \lam_i$ and $h_{i}$ and the search time  $T_{i}$ are constant over the study period, the unknown parameters $\lam$ and $h$ are not identifiable \citep{haines:16b}. For this reason, the model is deemed to be fragile and is indicated as such in the flow chart in Figure 1.
\vspace{2.5mm}

\noindent{\sc Binary:M}

The expression for the likelihood developed here for the Binary:M model  is based on the expansion introduced by \citet{haines:16b} for the Royle-Nichols model but is more nuanced. Consider first a single site with multiple visits, that is $J > 1$. For clarity, the subscript $i$ relating to the site is suppressed and the binary responses are denoted $y_j, j=1, \ldots, J$.
Suppose now that occasions are arranged so that the species is detected on the first $J_1$ occasions with search times flagged as $T_j^{(1)}$ and not detected on the remaining  $J - J_1$ occasions with search times flagged as $T_j^{(0)}$. The pmf of the responses summed over the number of individuals at the site, $n$, by invoking the Poisson mixing distribution is thus given by   
$$
g(y_1, \ldots, y_J) =  \sum_{n=0}^{\infty} \Big ( \prod_{j=1}^{J_1} (1-e^{-n h_j T_j^{(1)}}) \prod_{j=J_1+1}^{J} e^{-n h_j T_j^{(0)}} \Big ) \frac{\lam^n e^{-\lam}}{n!}.
$$
Consider now expanding the leading product term within the summation  as 
$$
 1 - \sum_{j=1}^{J_1} e^{-n h_j T_j^{(1)}} + \sum_{j=1}^{J_1-1} \sum_{k=j+1}^{J_1}  e^{-n (h_j T_j^{(1)}+ h_k T_k^{(1)})} -  \ldots  \pm e^{-n \sum_{j=1}^{J_1} h_j T_j^{(1)}}. 
$$
It then follows that the infinite sum embedded in the pmf $g(y_1, \ldots, y_J)$ can be evaluated explicitly. Specifically, setting  $W_{j}^{(1)} = h_j  T_j^{(1)}, \; W_{jk}^{(1)} = h_j  T_j^{(1)}+h_k T_k^{(1)}, \ldots,$ $W_A^{(1)}= \ds \sum_{j=1}^{J_1} h_j T_j^{(1)}$ and  $W_A^{(0)}=\ds \sum_{j=J_1+1}^{J} h_j T_j^{(0)}$  yields 

\begin{eqnarray*}
g(y_1, \ldots, y_J)  
& = &   \exp[-\lam(1- e^{- W_A^{(0)}})] \Big[ 1 - \sum_{j=1}^{J_1} \exp(\lam  e^{-W_j^{(1)} })   \Big.\\
& & \hspace{20mm} +\sum_{j=1}^{J_1-1} \sum_{k=j+1}^{J_1} \exp( \lam e^{- W_{jk}^{(1)}})- \Big.  \ldots  \pm \exp(\lam e^{- W_A^{(1)}}) \Big] .  
\end{eqnarray*}
If $J_1=0$, no individuals are detected on all occasions and the trailing term in this expression reduces, correctly, to one,  to give
\begin{equation}
P(y_1 = 0, \ldots, y_J=0) =\exp[-\lam (1 - e^{-\sum_{j=1}^J h_j T_j})].
\label{proby0m}
\end{equation} 
Overall therefore, the likelihood of the parameters over $R$ sites, arranged so that the species is detected on at least one occasion at the first $R_1$ sites and not detected at the remaining $R-R_1$ sites, can be expressed, in an obvious notation, as 
\begin{eqnarray*}
L_{Binary:M}(\btheta, \bphi;\by)  & = & \prod_{i=1}^{R_1} \Big[ 1 - \sum_{j=1}^{J_{i,1}} \exp(\lam_i  e^{-W_{i,j}^{(1)} }) + \sum_{j=1}^{{J_{i,1}}-1} \sum_{k=j+1}^{J_{i,1}} \exp(\lam_i e^{- W_{i,jk}^{(1)}}) -   \Big. \\
& & \hspace{17mm} \ldots  \Big.\pm \exp(\lam_i e^{- W_{i,A}^{(1)}}) \Big] 
\times \prod_{i=1}^R  \exp[-\lam_i(1- e^{- W_{i,A}^{(0)}})],  
\end{eqnarray*}
\noindent where $\by$ is the vector of binary responses, arranged conformably.
\bigskip

{\bf 3.2 Models for Time-to-First-Detection}
\label{sec:M2}

\noindent {\sc BinaryT1:S}

The time of arrival of an individual at site $i$ is exponentially distributed with parameter $h_i$ and, for $n_i$ individuals acting independently, it follows  from a standard result in order statistics that the time of the first arrival of the species is exponentially distributed with parameter $n_i h_i, i=1, \ldots, R$. Thus, for a search time $T_i$, the time-to-first-detection $t_{i(1)}$ for a binary response $y_i=1$  is a right-censored random variable with  pdf $n_i h_i e^{-n_i h_i t_{i(1)}}/(1-e^{-n_i h_i T_i})$. Then, since $P(y_i=1|n_i) = 1-e^{-n_i h_i T_i}$ and $P(y_i=0|n_i) = e^{-n_i h_i T_i}$, it follows that the pdf of the time-to-first-detection, given $n_i$, can be expressed as 
$
\big\{ n_i h_i e^{-n_i h_i t_{i(1)}}\big\}^{y_i}  \; 
\big\{e^{-n_i h_i T_i}\big\}^{1-y_i} 
 \mbox{~for~}  y_i = 0 \mbox{~or~} 1
$.
Thus, the pdf of $t_{i(1)}$ for $y=1$ can be derived as  
\begin{eqnarray*}
g(t_{i(1)}, y|y_{i}=1)  &= & \sum_{n_i=1}^{\infty}  n_i  h_i e^{- n_i h_i t_{i(1)}} \frac{\lam_i^{n_i} e^{-\lam_i}}{n_i!} \\
& = & h_i \lam_i \exp [ -\lam_i (1 - e^{-h_i t_{i(1)}}) - h_i t_{i(1))} ],\\ 
&& \hspace{10mm}  0 < t_{i(1)} < t_{i(2)}< \ldots < T_i \mbox{~or~} 0 < t_{i(1)} < T_i < t_{i(2)},
\end{eqnarray*}
where $t_{i(2)}$ is the second order statistic of the arrival times. Otherwise, for $y_i=0$, the time-to-first-detection is not recorded and, from expression (\ref{proby0s}), $P(y_i=0) =  \exp[-\lam_i ( 1- e^{-h_i T_i} )]$.
Consider now $R$ sites arranged so that a time-to-first-detection is recorded at the first $R_t$ sites and not recorded at the remaining $R-R_t$ sites. Then the  likelihood of the parameters is given, in  an obvious notation, by
\begin{eqnarray*}
L_{BinaryT1:S}(\btheta, \bphi; \by, \bt_{(1)}) & = & \prod_{i=1}^{R_t} h_i \lam_i \exp [ -\lam_i (1 - e^{-h_i t_{i(1)}}) - h_i t_{i(1)} ] \\
& & \hspace{20mm} \times \prod_{i=R_t+1}^{R} \exp[-\lam_i ( 1- e^{-h_i T_i} )], 
\end{eqnarray*}
where $\by$ is the vector of binary responses and $\bt_{(1)}$ the vector of the $R_t$  times-to-first-detection, arranged conformably.
\vspace{2.5mm}

\noindent {\sc BinaryT1:M}

Consider now a single site on $J>1$ occasions, with the subscript $i$ suppressed and the search times denoted $T_j, j=1, \ldots, J$.  Suppose that the occasions are arranged so that the first $J_t$ have $y_j=1$ and an observed time-to-first-detection  $t_{j(1)} < T_j, j=1, \ldots, J_t$, and that no detections are recorded on the remaining $J - J_t$ visits.
For $J_t > 0$, the number of individuals at the site, $n$, is necessarily greater than zero and the conditional joint pdf of the times-to-first-detection can be derived as 
\begin{eqnarray*}
g(\bt_{J_t(1)} | J_t > 0) & = & \sum_{n=1}^{\infty} \left [ \prod_{j=1}^{J_t} n h_j e^{- n h_j t_{j(1)}} \prod_{j=J_t+1}^J e^{-n h_j T_j} \times \frac{\lam^n e^{-\lam}}{n!} \right ]\\ \nonumber
& = & \ds \Big( \prod_{j=1}^{J_t} h_j \Big) \; e^{-\lam}  \sum_{n=0}^{\infty}  n^{J_t} \frac{(\lam e^{-W_t})^n} {n!}  \\ \nonumber
& = & \Big( \prod_{j=1}^{J_t} h_j \Big)  \exp[-\lam (1- e^{-W_t})]  \sum_{k=0}^{J_t} (\lam e^{- W_t})^k  S(J_t,k),
\end{eqnarray*}
where $W_t=\sum_{j=1}^{J_t} h_j t_{j(1)}+\sum_{j=J_t+1}^J h_j T_j$ and $\bt_{J_t(1)}$ is the vector of the $J_t$ times-to-first-detection. Note that the trailing summation  in the second line of the pdf is  proportional to the $J_t$-th moment of the Poisson distribution with parameter $\lam e^{-W_t}$ and, in the third line, $S(J_t,k)$ denotes a Stirling number of the second kind \citep[p. 162]{jkk:05}. 
For $J_t=0$, no individuals are detected over all $J$ occasions, that is $y_j=0, j=1,\ldots, J$, and the attendant probability is given by expression (\ref{proby0m}), that is by $\exp[-\lam (1-e^{-\sum_{j=1}^{J} h_j  T_j})]$.
Consider now $R$ sites,  arranged so that $J_{i,t} > 0$ at the first $R_t$ sites and $J_{i,t}=0$ at the remaining $R-R_t$ sites.   
Then  the likelihood of the parameters is given, in an obvious notation, by  
\begin{eqnarray*}
L_{BinaryT1:M}(\btheta, \bphi; \by, \bt_{(1)})& = & \ds \prod_{i=1}^{R_t} \bigg\{ \Big( \prod_{j=1}^{J_{i,t}}    h_{ij} \Big )  \exp \left [-\lam_i (1-e^{-W_{i,t}} ) \right ] \\
& & \hspace{15mm} \times  \ds \sum_{k=0}^{J_{i,t}} (\lam_i e^{-W_{i,t}})^k  S(J_{i,t},k) \bigg\}  \\  
& & \hspace{20mm} \times  \ds \prod_{i=R_t+1}^{R}  \exp[-\lam_i (1-e^{-\sum_{j=1}^{J} h_{ij}  T_{ij}})], 
\end{eqnarray*} 
where  $\by$ denotes the vector of binary responses and $\bt_{(1)}$ the vector of times-to-first-detection, arranged conformably, and $W_{i,t} = \sum_{j=1}^{J_{i,t}} h_{ij} t_{ij(1)}+ \sum_{J_{i,t}+1}^J h_{ij} T_{ij}$.
\bigskip

{\bf 3.3 Models for Count Data}
\label{sec:M3}

The Count:S and Count:M models  involve counts recorded over $R$ sites on each of  $J$ occasions and correspond to the binomial $N$-mixture model introduced in a somewhat different context by \citet{royle:04}. The results presented here are based on, but extend, those given for the latter  model in the paper by \citet{haines:16a}.
\vspace{2.5mm}

\noindent{\sc Count:S}

Suppose that  $J=1$ and that $y_i$ denotes the number of individuals counted  at site $i$ over the search time $T_i$, $i=1, \ldots, R$. Then the counts follow a thinned Poisson distribution  \citep{wadley:49} and the likelihood of the parameters is given by
$$
L_{Count:S}(\btheta, \bphi; \by) = \prod_{i=1}^R \frac{[\lam_i (1- e^{-h_i T_i})]^{y_i} \exp
[-\lam_i (1- e^{-h_i T_i}) ]}{y_i!},
$$
where $\by =(y_1, \ldots, y_R)$. Note that, if the parameters $\lam_i$ and $h_i$ and the search times $T_i$ are constant over all sites, that is $\lam_i=\lam, h_i=h$ and $T_i=T$, the unknown parameters $\lam$ and $h$ are not identifiable \citep{haines:16a}. The model is therefore fragile and is indicated as such in the flow chart in Figure 1.
\newpage
\noindent {\sc Count:M} 

Consider now a single site with $J > 1$ and the subscript $i$ suppressed. Suppose that, on the $j$th occasion, there are  $y_{j}$ detections, each associated with a probability of detection of an individual of $p_j, j=1, \ldots, J$. Then, if the maximum count at the site over all occasions, denoted $y_{M}=\underset{j=1,\ldots, J}{\max} ~y_j$, is greater than zero, the derivation in \citet{haines:16a} for a constant probability of  detection of an individual  can be modified straightforwardly to yield the joint pmf of the vector of counts given $y_M > 0$ explicitly as 
\begin{eqnarray*}
g(\by_J | y_M > 0) &=& e^{-\lam} \bigg[ \prod_{j=1}^J \left  (\frac{p_{j}}{1-p_{j}} \right )^{y_{j}} \bigg] \frac{\left [\lam \prod_{j=1}^{J} (1-p_j)\right ]^{y_{M}}}{y_{M}!} \; \prod_{j=1}^{J-1} \binom{y_{M}}{y_{j}} \nonumber \\ 
& &   \hspace{25mm} \times \hg{J-1}{J-1} 
\left( \left . \begin{array}{c}
(y_{M}+1)\\ 
(y_{M}-y_{ij}+1)
\end{array}
\right | \lam \prod_{j=1}^{J} (1-p_j) \right),
\end{eqnarray*}
where $\by_J=(y_1, \ldots, y_J)$ and the trailing term represents the generalised hypergeometric function
$$
\hg{J-1}{J-1} \left ( 
\left . \begin{array}{ccccc}
y_{M}+1, & \ldots, & y_{M}+1, & \ldots, & y_{M}+1\\ 
y_{M}-y_{1}+1,& \ldots, & y_{M}-y_{j}+1, & \ldots, & y_{M}-y_{J-1}+1
\end{array}
\right | \lam  \prod_{j=1}^J (1-p_j)  
\right).
$$
Otherwise, if all $J$ counts at a site are zero, that is $y_M=0$, the probability of the vector of counts is $\exp[-\lam \prod_{j=1}^J (1-p_j)]$ and is in accord with setting $y_M=0$ in the conditional density given $y_M>0$. In the present context, $p_{j}=1-e^{-h_{j} T_{j}}, j=1, \ldots, J$. Thus, if the sites are arranged so that $y_{i,M} > 0$ at the first $R_1$ sites,  and the counts at the remaining $R-R_1$ sites are all zero, the likelihood of the parameters, with irrelevant terms omitted, is given by
\begin{eqnarray*}
L_{Count:M}(\btheta, \bphi;\by) & = &   \prod_{i=1}^{R_1} \bigg\{ \; e^{-\lam_i}  \;\prod_{j=1}^J \left  (\frac{1-e^{-h_{ij} T_{ij}}}{e^{-h_{ij} T_{ij}}} \right )^{y_{ij}} \; \left ( \lam_i  e^{- \sum_{j=1}^{J} h_{ij} T_{ij}} \right )^{y_{i,M}} \\
& & \hspace{10mm} \times    \;  \hg{J-1}{J-1} 
\bigg( 
 \begin{array}{c}
(y_{i,M}+1)\\ 
(y_{i,M}-y_{ij}+1)
\end{array}
\bigg| \lam_i  e^{-\sum_{j=1}^{J} h_{ij} T_{ij}}) \bigg) \bigg\}                    \\
& & \hspace{20mm} \times \prod_{i=R_1+1}^R \exp[-\lam_i (1 - \exp^{-\sum_{j=1}^{J} h_{ij} T_{ij}})],
\end{eqnarray*}
where $\by$ is the vector of counts, arranged conformably.
Note that, for clarity of exposition and ease of computation, the  products over the sites with $y_{i,M} >0$ are presented separately from those with $y_{i,M}=0$. 
\bigskip
\newpage

{\bf 3.4  Models for Times-to-Each-Detection}
\label{sec:M4}

\noindent {\sc CountT:S}

Consider a single site with $J =1$ and the site index $i$ suppressed. Suppose first that $y > 0$ individuals are detected over a search time $T$ and that  $(n-y)$ individuals remain undetected, where $n$ is the unknown number of individuals at the site.  The  distribution of the detection times $t_d, d=1, \ldots, y$, given $y > 0$, can be derived by somewhat detailed arguments based on the theory of order statistics  and is given by 
$$
\ds g(\bt_d|y>0) =  \frac{h^{y} \; e^{-h \sum_{d=1}^y t_{d}}}{(1-e^{-h T})^y} \mbox{~~for~} \bt_d=(t_1, \ldots, t_y).
$$   
Crucially, the density depends on the parameter $h$ but not on the number of individuals at the site, $n$, and hence on the abundance parameter  $\lam$. This observation resonates with the fact that the times-to-each-detection, given $y>0$, depend only on the exponential distribution with parameter $h$ and that information on $\lam$, together with $h$, is embedded in the count $y$. 
In addition, and in accord with intuition, the times-to-each-detection only appear  in the sum, $\sum_{d=1}^y t_{d}$.

The joint probability density of the times $\bt_d$ and $y$, for $y > 0$, can now be derived by recalling 
that the count $y$ follows a thinned Poisson distribution with parameter $\lam (1-e^{-h T})$. The density is therefore given by
\begin{eqnarray*}
g(\bt_d, y| y>0) 
& = & \frac{h^y \lam^y}{y!} e^{ -h \sum_{d=1}^y t_{d} } \; \exp[-\lam(1-e^{- h T})] \mbox{~for~} 0<t_d <T, d=1, \ldots, y, 
\end{eqnarray*}  
and, for $y=0$, reduces, correctly, to $P(y=0) =  \exp[-\lam(1-e^{-h T})]$.
It now follows that, for $R$ sites arranged so that $y > 0$ at the first $R_t$ sites and $y=0$ at the remaining $R-R_t$ sites, the likelihood  of the parameters can be expressed in a simplified form as 
\begin{eqnarray*}
L_{CountT:S}(\btheta, \bphi; \by, \bt) & = & \prod_{i=1}^{R_t} \bigg[\frac{h_i^{y_i}  e^{- h_i \sum_{d=1}^{y_i} t_{i,d}}}{(1-e^{-h_i T_i})^{y_i}} \bigg]
\times 
L_{Count:S}(\btheta,\bphi; \by),
\end{eqnarray*}
where the trailing term is the likelihood for the counts alone, that is for model Count:S. The form of the likelihood, $L_{CountT:S}(\btheta, \bphi; \by, \bt)$, therefore reveals that more information about the  hazard can be gained by recording counts and times-to-each-detection than by recording counts alone. This observation  in turn impacts on, and should improve, the precision with which the parameters embedded in $\btheta$ are estimated.  
A complete derivation of the results is provided in the supplementary materials.
\vspace{2.5mm}

\noindent{\sc CountT:M}

Consider now a single site with more than one visit, that is $J > 1$, and the site index $i$ suppressed. The responses are  counts of individuals at each visit, $y_j$, with attendant search times $T_j$ and, if $y_j >0$, the times-to-each-detection, denoted $\bt_j=(t_{j1},\ldots, t_{jy_j})$, $j=1, \dots, J$. Suppose now that there are $J_t$ visits on which at least one $y_j >0$ and $J-J_t$ visits with no detections and that the visits are arranged so that the $J_t$ sites appear first. Then, since visits to a site are conducted independently, it follows that the pdf of the times-to-each-detection over the $J_t$ visits, given the counts, can be expressed as 
$$
g(\bt_1, \ldots, \bt_{J_t}|J_t > 0) = \ds \prod_{j=1}^{J_t} \frac{ \; h_{j}^{y_j} \;
e^{- h_{j} \sum_{d=1}^{y_{j}} t_{jd}} }
{\ds \left ( 1-e^{-h_{j} T_{j}} \right )^{y_{j}}}.
$$ 
Clearly, for the $J-J_t$ occasions on which $y_j=0$, only the count itself is recorded. 
The likelihood of the parameters with the sites arranged so that $R_t$ sites with a maximum response greater than $0$ appear first now follows and can be expressed succinctly as 
\begin{eqnarray*}
L_{CountT:M}( \btheta, \bphi;\by, \bt) & = &  \prod_{i=1}^{R_t} \prod_{j=1}^{J_{i,t}} \bigg[ \frac{h_{ij}^{y_{ij}} \; e^{-h_{ij} \sum_{d=1}^{y_{ij}} t_{i,jd}}}{\ds \left ( 1-e^{-h_{ij} T_{ij}} \right )^{y_{ij}}} \bigg] \times L_{Count:M}(\btheta, \bphi; \by),
\end{eqnarray*}
where the trailing term is the likelihood of the parameters for the counts only, that is for model Count:M. 
Thus, more information about the hazard, $h$,  is gained by recording counts and times-to-each-detection  than by recording counts alone. 
A complete derivation of these results is also included in  the supplementary materials.
\bigskip

{\bf 3.5  Count and time-to-first-detection}
\label{sec:M5}

\noindent{\sc CountT1:S and CountT1:M}

The broad structure of the derivations of the likelihoods for the CountT1 models follows that for the CountT and is therefore outlined briefly here.
Consider first a single site with $J=1$ and, for clarity, with the index $i$ suppressed. Suppose that there are $y$ individuals detected in the search time $T$ and that, if $y >0$, the  time-to-first-detection $t_{(1)}$ is also recorded. It then follows that the pdf of $t_{(1)}$, conditional on $y >0$, is given by 
$$
g(t_{(1)}| y>0) =
y \; h \; e^{-h_i t_{i(1)}} \; \frac{[e^{-h t_{(1)}} -e^{- h T} ]^{y-1}}{(1- e^{-h T})^{y}}
$$
and does not depend on the number of individuals at the site, $n$. It is interesting to note that this joint density with $y=1$ is in accord with that derived for the model CountT:S with $y=1$ and $t_{(1)} < T < t_{(2)}$ but, crucially, differs from that for the model BinaryT1:S which models $t_{(1)}$ alone. 

The derivation for model CountT1:S now follows immediately from that for model CountT:S, and the likelihood of the parameters, with $y_i > 0$ at the first $R_t$ sites and  $y_i = 0$ at the remaining  $R - R_t$ sites, can be expressed as 
\begin{eqnarray*}
L_{CountT1:S}(\btheta, \bphi; \by, \bt_{(1)}) & = & \prod_{i=1}^{R_t} \Big\{  \; \frac{h_i e^{-h_i t_{i(1)}} [e^{-h_i t_{i(1)}} -e^{- h_i T_i} ]^{y_i-1}}{(1- e^{-h_i T_i})^{y_i}}  \Big\} \\
& & \hspace{40mm} \times L_{Count:S}(\btheta, \bphi; \by),
\end{eqnarray*}
where $\by$ and $\bt_{(1)}$ are vectors representing the counts and the times-to-first-detection, arranged conformably.
Similarly, the derivation for model CountT1:M follows  that for model CountT:M and the likelihood of the parameters is given, in a compatible notation, by 
\begin{eqnarray*}
L_{CountT1:M}( \btheta, \bphi;\by, \bt) & = &  \prod_{i=1}^{R_t} \bigg\{  \prod_{j=1}^{J_{i,t}}  \frac{h_{ij}^{y_{ij}} \; e^{-h_{ij} t_{ij(1)}} (e^{-h_{ij} t_{ij(1)}} -e^{- h_{ij} T_{ij}} )^{y_{ij}-1}}{(1- e^{-h_{ij} T_{ij}})^{y_{ij}}} \bigg\} \\
& & \hspace{40mm} \times L_{Count:M}(\btheta, \bphi; \by)
\end{eqnarray*}
where $\by$  and $\bt_{(1)}$ are vectors representing the counts and times-to-first-detection, arranged conformably. For both of the CountT1 models, information is clearly gained by recording times-to-first-detection and counts as opposed to counts alone. However, since only a single time is recorded for these models, it is tempting to infer that the gain in information is less than that for the CountT models.
\bigskip
\bigskip

{\bf 4. A Poisson Process}
\label{sec:hpp} 

Suppose that an individual may be detected and thereby counted more than once during a single visit to a site. This setting can be modelled by again taking the mixing distribution  to be Poisson with parameter $\lam_i(\bphi), i=1, \ldots, R$. However, the observation process is now formulated on the premise that a single individual is detected at a site and on a single occasion according to a homogeneous Poisson process \citep{garml:12}. The five models introduced in Section 2 are considered within this context and are identified by the notation PBinary through PCountT1.  The rate parameter of the Poisson process at site $i$ on occasion $j$ is denoted by $\g_{ij} (\btheta_P)$, with $\btheta_P$ an unknown parameter, and is denoted compactly as $\g_{ij}, i=1, \ldots, R, j=1, \ldots, J$. Then, by the principle of superposition, individuals arrive at a site on a single visit according to a homogeneous Poisson process with parameter $n_i \g_{ij}$, where $n_i$ is the unknown number at the site. This result is used throughout the derivations which now follow. Otherwise, the same notation as that introduced for the binomial count approach is adopted. The difference  between the two observational processes, the binomial count and the Poisson, is highlighted with a question and the appropriate models are clearly delineated in the flow chart in Figure 1.
\bigskip

{\bf 4.1 Poisson-based models for detection/non-detection and time-to-first-detection}
\label{sec:M12P}
\noindent{\sc PBinary and PBinaryT1}

Consider a site $i$ on an occasion $j$, with the number of individuals at the site $n_i$. Then, for the  PBinary model, the probability of no detections given $n_i$ is $\exp^{-n_i \g_{ij} T}$ and, for the model PBinaryT1 with a binary response or count $y_{ij}> 0$, the time-to-first-detection given $n_i$ is exponentially distributed with parameter $n_i \g_{ij}$. These expressions are identical in form to those for the Binary and BinaryT1 models  obtained by invoking the binomial count approach and the likelihoods of the parameters, with the exponential parameter $h_{ij}$ replaced by the rate $\g_{ij}$, follow accordingly.  It should be emphasised however that the results for the two observation processes, that is the binomial count approach and the homogeneous Poisson process, while the same in formulation, are conceptually very different.  The connection between the Binary and the PBinary models, albeit algebraic, is illustrated clearly in the flow chart in Figure 1.
\bigskip

{\bf 4.2 Poisson-based models for counts}
\label{sec:M3P}

\noindent{\sc PCount:S}

An explicit expression for the likelihood of the unknown parameters for the PCount:S model with a single visit to each site was derived by \cite{garml:12} within the context of distance rather than time. Specifically, the authors modelled the number of paw marks along a transect, based on the assumption that animals could cross the transect more than once, by invoking a homogeneous Poisson process with the counts therefore following a Poisson distribution. In the present case, the number of detections at site $i$, denoted $y_i$, over a search time $T_i$ is similarly Poisson distributed with parameter $n \g_i T_i$, where $i=1, \ldots, R$. For $y_i > 0$, the pmf of the count, given $n$, follows as 
$$
g(y_i |y_i > 0, n) =  \sum_{n=0}^{\infty} \Big[ \frac{e^{-n \g_i T_i} (n \g_i T_i)^{y_i}}{y_i!} \times \frac{\lam_i^n e^{-\lam_i} }{n!} \Big]. 
$$
The count therefore follows a Neyman Type A distribution, that is a Poisson mixture of Poisson distributions defined by Poisson($\btheta_P$)$\underset{\fns{\btheta_P}/\g T}{\bigwedge}$ Poisson($\lam$) \citep[p.403]{jkk:05}. In addition, the probability that  $y_i=0$ follows from expression (\ref{proby0s}). Thus,  based on the results of \cite{garml:12}, the likelihood of the parameters $\bphi$ and $\btheta_P$ at the vector of counts $\by = (y_1, \ldots, y_R)$,  with $y_i >0$ at the first $R_1$ sites and $y_i=0$ at the remaining $R-R_1$ sites, is given. by
\begin{eqnarray*}
L_{PCount:S}(\btheta_P, \bphi | \by) & = & \prod_{i=1}^{R_1} \bigg\{ \frac{(\g_i T_i)^{y_i}}{y_i!}  \exp[-\lam_i (1- e^{-\g_i T_i})] \; \ds  \sum_{k=0}^{y_i} (\lam_i e^{-\g_i T_i})^k S(y_i,k) 
 \bigg\} \\
& & \hspace{30mm} \times  \prod_{i=R_1+1}^R
\exp[-\lam_i (1- e^{-\g_i T_i})],
\end{eqnarray*} 
where $S(y_i,k)$ is again a Stirling number of the second kind. The likelihood can be represented more succinctly as
$$
L_{PCount:S}(\btheta_P, \bphi | \by) = \prod_{i=1}^{R_1} \bigg\{ \frac{(\g_i T_i)^{y_i}}{y_i!} \; \ds  \sum_{k=0}^{y_i} (\lam_i e^{-\g_i T_i})^k S(y_i,k) 
 \bigg\}  \times  \prod_{i=1}^R
\exp[-\lam_i (1- e^{-\g_i T_i})],
$$
a form which is computationally efficient.
\vspace{2.5mm}

\noindent {\sc PCount:M}

The derivation of the likelihood  of the parameters for the model PCount with multiple visits to each site is more nuanced than that for the single visit model PCount:S. Specifically, the distribution of the counts is no longer Neyman A and the results of  \citet{garml:12} do not apply. The pmf of the counts and the likelihood for the model PCount:M are therefore developed here from first principles. Thus, consider a single site on $J >1$ occasions with index $i$ suppressed and with rate parameters $\g_j$ and search times $T_j$, where $j=1, \ldots, J$. The counts at each visit comprise a vector $\by_J = (y_1, \ldots, y_J)$, with the maximum count at the site denoted $y_M$. The pmf of $\by_J$, given $y_M >0$, can now be derived straightforwardly. Thus, for the  visits arranged so that $y_j > 0$ for the first $J_y$ visits and $y_j=0$ for the remaining $J-J_y$ visits, it follows that
\begin{eqnarray*}
g(\by_J| y_M > 0) & = &  \sum_{n=1}^{\infty} \;\bigg[ \prod_{j=1}^{J_y}  \frac{(n \g_j T_j)^{y_j} e^{-n \g_j T_j} }{y_j!} \prod_{j=J_y+1}^J  e^{-n \g_j T_j} \times \frac{ \lam^n e^{-\lam}}{n!} \bigg] \\
 & = & \ e^{-\lam} \prod_{j=1}^{J_y} \frac{(\g_j T_j)^{y_j}}{y_j!}  \times 
\sum_{n=0}^{\infty}  n^{y_{+}}\frac{(\lam e^{- \sum_{j=1}^J \g_j T_j})^n}{n!}  \\
& = & \prod_{j=1}^{J_y} \frac{(\g_j T_{j})^{y_j}}{y_j!}  \exp[-\lam (1-e^{-\sum_{j=1}^J \g_j T_j})] \ds \sum_{k=0}^{y_{+}} (\lam e^{-\sum_{j=1}^J \g_j T_j})^k S(y_{+} ,k), 
\end{eqnarray*}
where $y_+$ denotes the sum $\sum_{j=1}^{J_y} y_j$. Note that the trailing term in the expression for this density is proportional to the $y_{+}$th moment of the Poisson distribution with parameter $\lam e^{-\sum_{j=1}^J \g_j T_j}$. For a site with all counts zero, that is $y_M=0$, the pmf  of the counts is given, following expression (\ref{proby0m}), by $\exp[-\lam (1-e^{-\sum_{j=1}^J \g_j T_j})]$. Thus, for $R$ sites arranged so that $y_{i,M} > 0$ for the first $R_1$ sites and $y_{i,M}=0$ for the remaining $R-R_1$ sites, the likelihood  of the parameters can be expressed as 
\begin{eqnarray*}
L_{PCount:M}(\by| \btheta_P, \bphi) & = & \prod_{i=1}^{R_1} \; \bigg\{
\prod_{j=1}^{J_{i,y}} \frac{(\g_{ij} T_{ij})^{y_{ij}}}{y_{ij}!} \;  \; \ds \sum_{k=0}^{y_{i+}} (\lam_i e^{-\sum_{j=1}^J \g_{ij} T_{ij}})^k S(y_{i+},k)   \bigg\}\\
& & \hspace{35mm}
 \times \prod_{i=1}^R \exp[-\lam_i (1-e^{-\sum_{j=1}^J \g_{ij} T_{ij}})], 
\end{eqnarray*}
where $\by$ is the vector of counts, arranged conformably, and $y_{i+}=  \sum_{j=1}^{J{i,y}} y_{ij}, i=1, \ldots, R$.
\bigskip

{\bf 4.3 Poisson models for times-to-each detection and counts with time-to-first-detection.}
\label{sec:M45P} 

\noindent{\sc PCountT and PCountT1}

Results for the PCountT and PCountT1 models are  predicated on the complete randomness of a homogeneous Poisson process.
Specifically, consider a single visit to a single site,  with the indices $i$ and $j$ suppressed and with the count $y$ over a search time $T$ greater than zero. Then, for the model PCountT, given that $y$ detections are recorded over the search time, the times-to-each-detection have the same joint distribution  as the order statistics of a random sample taken from the uniform distribution over the interval $(0,T)$. The joint density of these times, conditional on $y>0$, is therefore given by 
$$
g(\bt|y > 0) = \frac{y!}{T^y} \mbox{~~for~} 0 < t_1 \ldots  t_y < T,
$$
where $\bt$ denotes the vector of detection times, $(t_1, \ldots, t_y)$ \citep[Theorem 2.3.1]{ross:96}. In addition, for the  PCountT1 model, the time-to-first-detection over the search time $T$, given the count $y$,  has pdf 
$$
g(t_{(1)} |y >0) = \frac{y}{T} \left (1-\frac{t_{(1)}}{T} \right )^{y-1}  \mbox{~~for~} 0 < t_{(1)} < \ldots < t_{(2)} < T \mbox{~or~}  t_{(1)} <  T < t_{(2)} \ldots  t_{(y)}
$$ 
or, in other words, $\ds \frac{t_{(1)}}{T}$ follows a beta distribution with parameters $1$ and $y$ \citep[p. 192]{adg:11}. 
Thus, for the PCountT and PCountT1 models based on a homogeneous Poisson process, it is clear that the conditional distributions of the times-to-each-detection, given the count, and of the time-to-first-detection, given the count, do not depend on the unknown parameters, $\btheta_P$ and $\bphi$, and that these results hold for all sites on all occasions. It therefore follows that the likelihoods of the parameters associated with the PCountT and PCountT1 models  on both single and multiple visits to the sites are, up to multiplying constants, the same as those for the models for counts alone, that is PCount:S and PCount:M. In other words, no further information on the parameters, $\btheta_P$ and $\bphi$, is obtained by recording detection times in addition to the counts. The results for the PCountT and PCountT1 models are therefore in sharp contrast to the corresponding models CountT and CountT1 for the observation process based on binomial counts. The equivalence between the parameter-based components of the likelihoods of the PCountT and PCountT1 models and those of the PCount model is illustrated in the flow chart in Figure 1. 
\bigskip
\bigskip

{\bf 4. Application}
\label{sec:disc}

The aim of this section is to introduce the models and the attendant likelihoods of Section 3 in a more accessible and easy-to-use form. To immediately simplify matters, the assumption that the exponential and rate parameters and the search times are site- and visit-dependent is relaxed. Specifically, the more straightforward setting in which abundance is taken to be either constant or site-dependent and the rate  parameters and search times to be constant is now considered.
An R package, {\tt nmtd}, which comprises functions for generating data and calculating the log-likelihood  for all the models of this setting has been developed and is provided as supporting material. The package has a help menu and includes simulated examples of the model fits for each function.  To reinforce the details of the code, formulae for the simplified likelihoods are included in the supplementary material. The use of the package is illustrated with data drawn from the Karoo dataset which is taken from the paper by  \citet{henry:20} and made available in the {\tt nmtd} package. The data comprise times-to-first-detection for the rufous-eared warbler ({\em Malcorus pectoralis}) and are analysed using the Binary:M and BinaryT:M models. Full details are available in the vignette which accompanies the R package.
\bigskip
\bigskip

{\bf 5. Discussion}
\label{sec:disc} 

The main results of the present study relate to the derivation of explicit expressions for the likelihoods of the parameters associated with $N$-mixture models for binary responses, counts and time-to-detection data. More specifically, the models considered are the detection/non-detection model of \citet{roylen:03}, extended to include time-to-first-detection data and the binomial $N$-mixture model of \citet{royle:04}, extended to include time-to-each-detection and time-to-first-detection data together with the counts.  A Poisson mixing distribution is assumed to hold  and  two observation processes are introduced. The first, termed the binomial count approach, accommodates settings in which an individual, if detected more than once on a single visit, is counted only once and the second, a homogeneous Poisson process, which models double counting. The specification of the parameters within the likelihood expressions is general, with abundance taken to be site-dependent and the exponential  and rate parameters and the search times, which determine the probability of detection of an individual, to be both site and occasion-dependent. The nature of the likelihoods associated with the $N$-mixture models is discussed throughout the text and and comparisons between the models, particularly in relation to the two observation processes, are also made. Arguably, the most crucial finding of the paper is that information on the parameters is gained by recording counts and times-to-detection as opposed to counts alone for the binomial count approach. This gain does not hold for the homogeneous Poisson process however. To conclude, the R package {\tt nmtd} complements the algebraic results and demonstrates the ease with which the fitting of the models to simulated and real-world data can be implemented computationally.

 There is considerable scope for further research, both in theory and in practice. From a theoretical perspective,  the assumptions of a Poisson mixing distribution and an exponential arrival time for the binomial count approach  made in the present study  are, admittedly,  somewhat restrictive.  \citet{strebel:21} introduced a range of  mixing and arrival time distributions for the first-time-to-detection $N$-mixture model and used an upper bound on the embedded infinite sum in order to compute the likelihoods. Developing exact likelihoods of the parameters for mixing distributions such as the negative binomial and arrival time distributions times such as the  Weibull would seem  to be  challenging. From a computational point-of view however there are considerable advantages to be gained from the models and the attendant exact likelihoods developed in this paper. The models can be readily implemented within both a frequentist and a Bayesian framework in programming languages such as R and Python and, to this end, the code embedded in the package {\tt nmtd} could be followed and extended. Introducing the exact forms of the likelihoods of the parameters into the routines, as opposed to placing an upper bound on the embedded sums within the likelihoods, should provide a considerable reduction in computer-time and also improve numerical efficiency. 
 
The main question arising within the present context relates to the extent to which time-to-detection data, together with binary response or count data, can enhance the precision with which the abundance and probability of detection of an individual are estimated.  Other  questions, {\em inter alia}, relate to the stability of the models, particularly with respect to outliers, to the impact of including covariates on that stability and to the design of surveys which provide satisfactorily precise parameter estimates and are, at the same time, cost effective.  Further studies are therefore needed to address these questions. In particular, the exact forms of the likelihoods of the parameters for the models of interest could be interrogated algebraically and by simulations with protocols based on the modelling framework of this paper. Work is currently in progress to examine the precision of the parameter estimates for a range of $N$-mixture  models using simulation and to align the results  with those derived  algebraically.
\bigskip
\bigskip

\noindent{\bf Supporting and Supplementary Material}

The package {\tt nmtd} and the accompanying vignette can be found on GitHub under \\ https://github.com/david-borchers/nmtd and are subject to {caveat utilitor: user beware}.

Supplementary materials can be obtained from the authors. The material comprises a complete derivation of the formulae for the likelihoods of the  CountT models and formulae for the models with only abundance taken to be covariate-dependent. 
\bigskip

{\bf Acknowledgements:} {The authors would like to thank Allan Clark and David Maphisa for many helpful discussions and Linda Haines would like to thank Iain MacDonald for sharing his insights into the theory of order statistics. Linda Haines and Res Altwegg would like to thank the University of Cape Town and the National Research Foundation (NRF) of South Africa for financial support. Any opinion, finding and conclusion or recommendation expressed in this material is that of the authors and the NRF does not accept liability in this regard.}
 
\vspace{10mm}
\bibliographystyle{chicago}
\bibliography{exactref}

\bigskip
\begin{figure}[htb!]
\vspace{-20mm}
\hspace{-14.5mm}\includegraphics[width=1.2\textwidth,height=1.2\textheight]{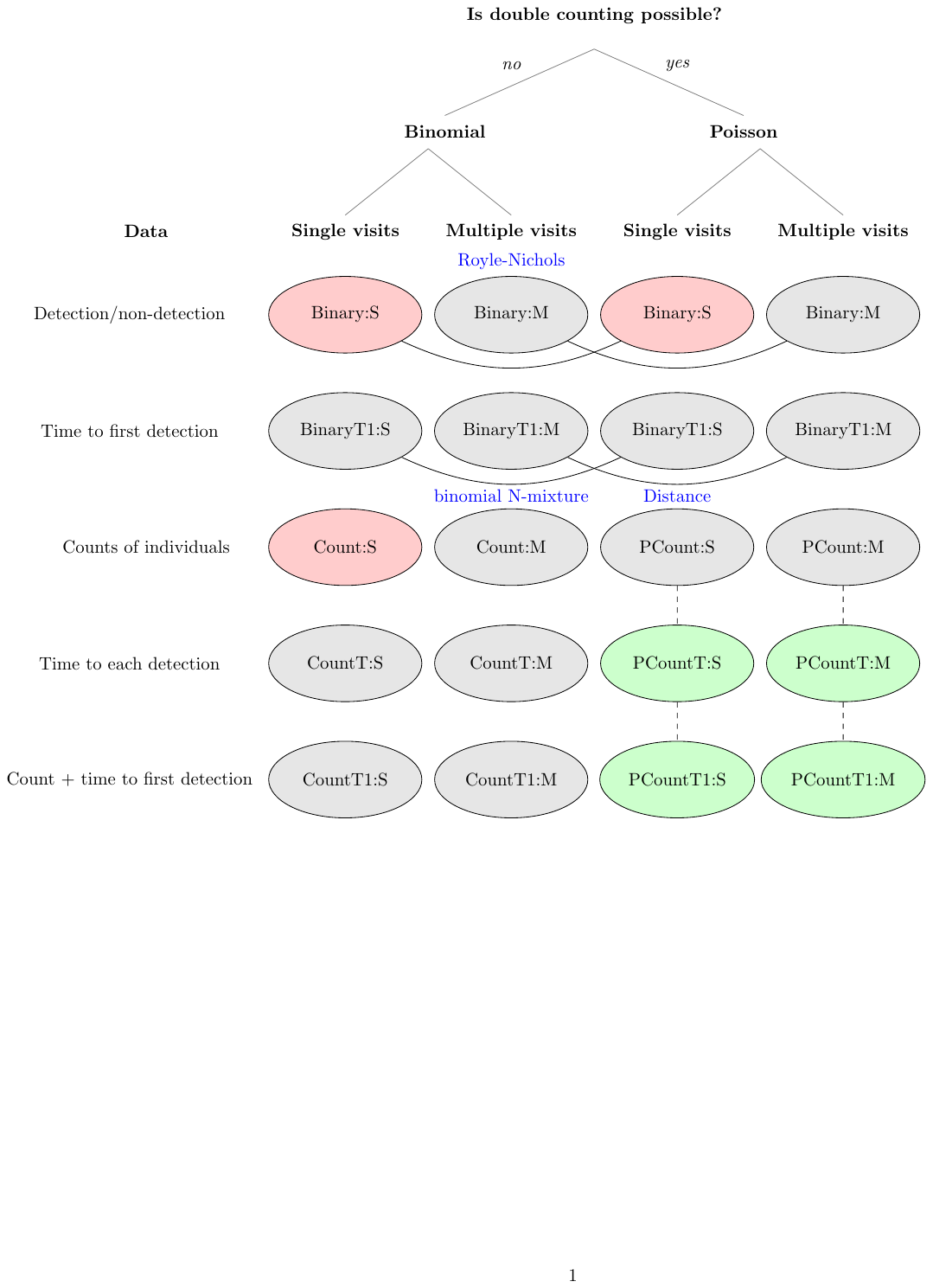}
\vspace{-100mm}

\caption{A Flow Diagram for the Models. Red ellipses indicate fragile models, that is models for which the parameters may not necessarily be identifiable. Ellipses linked by solid lines indicate that the models are identical. Green ellipses indicate models for which time is not informative and the attendant dashed lines emphasise that the models share the same parameter-dependent kernel of the likelihood as the parent PCount models. Models from the literature are indicated above the ellipses of the models in blue, with {\color{blue}{Distance}} referring to the paper by Guillera-Arroita et al. (2012).}
\label{fig1}
\end{figure} 

\end{document}